\documentstyle[aps,epsfig]{revtex}

\newcommand{\beq}{\begin{equation}}
\newcommand{\eeq}{\end{equation}}
\newcommand{\beqa}{\begin{eqnarray}}
\newcommand{\eeqa}{\end{eqnarray}}
\newcommand{\Psib}{\overline{\Psi}}

\newcommand{\Lcal}{{\cal L}}
\newcommand{\p}{\partial}
\begin{document}                
%
%
%
\title{Dirac-Brueckner Approach to Hyperon Interactions and Hypernuclei}
\author{H. Lenske, C.M. Keil, F. Hofmann, S. Briganti \\
Institut f\"ur Theoretische Physik, Universit\"at Gie\ss en\\
         Heinrich-Buff-Ring 16, D-35392 Gie\ss en, Germany}
%
\maketitle
\begin{abstract}
The density dependent relativistic hadron (DDRH) theory is
introduced as an effective field theory for nuclei and
hypernuclei. A Dirac-Brueckner approach to in-medium
nucleon-hyperon interactions is presented. Density dependent
meson-baryon vertices are determined from DBHF self-energies in
infinite matter. Scaling laws for hyperon vertices are derived
from a diagrammatic analysis of self-energies. In a local density
approximation the DB vertices are applied in relativistic DDRH
Hartree calculations to finite hypernuclei. $\Lambda$ single
particle spectra and spin-orbit splittings are described
reasonably well over the whole mass region.
\end{abstract}
\section{Introduction}               
\label{Intro}

Hypernuclear studies are the natural extension of isospin
dynamics in non-strange nuclei towards a more general theory of
flavor dynamics in a baryonic environment. Modern hypernuclear
theories are using non-relativistic and relativistic microscopic
descriptions \cite{Ba90,Ch89}. Since from a QCD point of view
hypernuclei as also isospin nuclei are deep in the
non-perturbative low energy-momentum regime, such a description
in terms of mesons and baryons should be adequate. Relativistic
mean-field (RMF) theories of Walecka-type \cite{SW86} have been
applied successfully \cite{Ru90,Gl92,Vr98} with empirically
adjusted meson-hyperon vertices. SU(3)$_f$-symmetric field
theories incorporating chirality \cite{Mu99,Pa99} or accounting
for the quark structure of hadrons \cite{Ts98} have been
formulated and applied to hypernuclei. In a SU(3)$_f$ approach
nucleon-hyperon and hyperon-hyperon interactions in free space
\cite{Re96,St99a,Ri99} and in a nuclear environment
\cite{St99b,Sc98} have been calculated.

In this contribution, hypernuclei are described in the Density
Dependent Relativistic Hadron (DDRH) theory. Introduced
originally as an effective field theory for isospin nuclei
\cite{LF95,FL95} it was extended to hypernuclei recently
\cite{KHL00}. In DDRH theory the in-medium modifications of
meson-baryon vertices are incorporated by functionals of the
fermion field operators. The functional dependence of the
vertices on density is derived from infinite matter
Dirac-Brueckner Hartree-Fock (DBHF)
calculations\cite{Jo98a,Jo98b}. For isospin nuclei, a practically
parameter-free model Lagrangian is obtained once a free space
interaction is chosen. Lorentz-invariance, thermodynamical
consistency and covariance of the field equations are retained.

The extension of DDRH theory to strange baryons is reviewed in
sect.\ref{sec:DensDepNucHyp}. As the central theoretical result
nucleon and hyperon vertices are found to be related by scaling
laws. A reduced model, appropriate for relativistic Hartree
calculations of single $\Lambda$ nuclei, is introduced. A
semi-microscopic derivation of the N$\Lambda$ interaction is
discussed where the $\sigma$ coupling is taken from a theoretical
N$\Lambda$ T-matrix \cite{Re96,Ha98} while the $\omega$ coupling
is determined empirically. DDRH mean-field results for
hypernuclei are presented in sect.\ref{sec:Spectroscopy} and
compared to RMF calculations. On a global scale spectroscopic
data are described satisfactorily well, including the reduced
spin-orbit splitting in $\Lambda$ nuclei. The paper closes with a
summary, conclusions and an outlook to work in progress in
sect.\ref{sec:summary}.

\section{Density dependent hadron field theory with hyperons}
\label{sec:DensDepNucHyp}

\subsection{The DDRH model Lagrangian}
\label{ssec:ModelLagr}

In hypernuclear models derived from a symmetry-broken SU(3)$_f$
Lagrangian \cite{ChengLi92} neither of the $0^-$ $\pi$ and K
meson fields contributes directly to a structure calculation,
except in the u-channel through antisymmetrization. From the
$1^-$ vector meson octet condensed isoscalar $\omega$ and
isovector $\rho$ meson fields will evolve. In a system with a
large fraction of hyperons also condensed octet $K^*$ and singlet
$\Phi$ mesons fields can appear. A shortcoming of a pure
SU(3)$_f$ approach is the missing of $0^+$ scalar mesons and,
hence, the absence of a binding mean-field. A satisfactory
description of the $0^+$ meson channels, e.g. in terms of
dynamical two-meson correlations \cite{Re96,Ha98}, is an unsolved
question.

In view of these problems we use an effective Lagrangian
including the degrees of freedom which are relevant for the
nuclear structure problem. This is achieved by extending the
original DDRH proton-neutron Lagrangian to the $1/2^+$ S=-1
($\Lambda, \Sigma^{\pm,0}$) and S=-2 ($\Xi^{-,0}$) hyperon
multiplets \cite{KHL00}. In the meson sector the isoscalar
$\sigma$, $\sigma_s$ ($\equiv$ scalar $s\overline{s}$
condensate), $\omega$ and $\phi$ meson, the isovector $\rho$
meson and the photon $\gamma$ are included. This leads to the
isospin-symmetric Lagrangian
\beqa\label{Lagrangian}
\Lcal &=& \Lcal_{B} + \Lcal_{M} + \Lcal_{int} \nonumber\\
\Lcal_{B} &=& \Psib_{F} \left[ i\gamma_\mu\p^\mu
                              - \hat{M} \right] \Psi_{F} \nonumber\\
\Lcal_{M} &=&\frac{1}{2} \sum_{i = \sigma, \sigma_s}
\left(\p_\mu\Phi_i\p^\mu\Phi_i - m_{\Phi_i}^2\Phi_i^2\right)
           - \frac{1}{2} \sum_{\kappa = \omega, \phi, \rho, \gamma}
             \left( \frac{1}{2} F^{(\kappa)^2} - m_\kappa^2 A^{(\kappa)^2}
\right) \label{eq:ModLagr} \\
\Lcal_{int} &=& \Psib_F \hat{\Gamma}_\sigma \Psi_F \Phi_\sigma -
\Psib_F \hat{\Gamma}_\omega \gamma_\mu \Psi_F A_{\omega}^\mu -
\frac{1}{2}\Psib_F \vec{\tau} \hat{\Gamma}_\rho \gamma_\mu \Psi_F
\vec{A}_{\rho}^\mu \nonumber\\
&&+ \Psib_F \hat{\Gamma}_{\sigma_s} \Psi_F \Phi_{\sigma_s} -
\Psib_F \hat{\Gamma}_\phi \gamma_\mu \Psi_F A_{\phi}^\mu - e
\Psib_F \hat{Q} \gamma_\mu \Psi_F A_{\gamma}^\mu, \quad .
\nonumber \eeqa
Here, $\Lcal_B$ and $\Lcal_M$ are the free baryonic and  mesonic
Lagrangians, respectively. Baryons are described by the flavor
spinor $\Psi_F$
\beq \Psi_F = \left( \Psi_N, \Psi_\Lambda, \Psi_\Sigma, \Psi_\Xi
\right)^{T} \label{eq:FlSp} \quad . \eeq
The diagonal matrix $\hat{M}$ contains the free-space baryon
masses and $\hat{Q}$ is the electric charge operator. The
meson-baryon interactions are contained in $\Lcal_{int}$. The
usual field strength tensor of either the vector mesons ($\kappa =
\omega, \phi, \rho$) or the photon ($\kappa = \gamma$) is denoted
by $F^{(\kappa)}_{\mu\nu}$. Contractions of the field strength
tensors are abbreviated as $F^2 = F_{\mu\nu} F^{\mu\nu}$ etc..

An important difference of the DDRH Lagrangian,
eq.(\ref{Lagrangian}), to standard RMF approaches
\cite{Ru90,Gl92} is the description of medium effects and
non-linearities in terms of density dependent meson-baryon vertex
functionals $\hat\Gamma_{\alpha B}=\hat\Gamma_{\alpha B}(\Psib_F
\Psi_F)$. The vertices are taken as Lorentz invariant functionals of the fermion
field operators $\Psi_F$ and since the baryon fields are treated
as quantum fields, even in the mean-field limit a well defined
class of quantum fluctuations with non-vanishing ground state
expectation values is taken into account \cite{FL95,KHL00}.
Dynamically, the vertices contribute to the Dirac equations as
rearrangement self-energies describing the static polarization of
the medium \cite{FL95,Ne82}. In bulk quantities, as for example
total binding energies, the DDRH rearrangement self-energies are
cancelled exactly but contribute to single particle quantities
like separation energies, wave functions and density matrices
\cite{FL95,KHL00}.

A solvable model is obtained in the Hartree mean-field
approximation. The operator-valued vertex functionals become
c-number functions of the baryon densities \cite{FL95,KHL00}. The
meson fields are obeying classical field equations, while the
baryons are treated as quantum fields by solving the Dirac
equation with static but density dependent self-energies,
including rearrangement contributions \cite{KHL00}. For single
$\Lambda$ nuclei the hidden-strangeness $\sigma_s$ and $\phi$
fields are neglected because they are of order $\cal{O}$(1/A)

\subsection{Dirac-Brueckner Approach to In-medium Hyperon Interactions}
\label{DB-int}

It is obvious that the structure of the vertex functionals must be
derived in a separate step. Provided that data are available the
required information could be obtained from phenomenology, as in
the approach of ref. \cite{Typel} for isospin nuclei. A
derivation on theoretical grounds has the advantage of providing
- at least in principle - a deeper insight into the structure of
interactions and, especially, the origin of medium-dependencies
and inter-relations between the flavour sectors. Hence, we follow
the original DDRH approach \cite{FL95} and derive the vertex
functionals from Dirac-Brueckner calculations. Full scale
SU(3)$_f$ DB calculations, however, are a considerable task whose
solution is still pending. But the success of non-relativistic
Brueckner calculations \cite{Vi98} clearly indicate the promising
potential of such a microscopic approach.

The extension of Dirac-Brueckner calculations to the SU(3)$_f$
multiplet have been outlined in \cite{KHL00}. Aiming at
applications of the DB-results in relativistic mean-field
calculations it is sufficient to have accurate knowledge on
baryon self-energies rather than on the full momentum structure of
in-medium interactions. A central result of \cite{KHL00} is the
observation that Dirac-Brueckner interactions can be expressed in
terms of medium-renormalized meson-exchange interactions with
density dependent vertex factors,
\beq\label{GammaB} \Gamma_{\alpha B}(k^B_F) \equiv g^B_\alpha
s^B_\alpha(k^B_F) \eeq
given by the bare coupling constants $g^B_\alpha$ and the density
dependent renormalization factor $s^B_\alpha(k^B_F)$ for baryons
of type B interacting through the exchange of mesons $\alpha$.

With our choice of momentum independent, global vertices, the
baryon self-energies are obtained as \cite{KHL00}
\beq\label{Gphi} \Sigma^B_\alpha(k|k^N_F,k^Y_F) = \Gamma_{\alpha
B}(k^B_F) \phi_\alpha(k|k^N_F,k^Y_F,\Gamma) \eeq
where $\Phi_\alpha$ denotes a condensed meson field. From this
equation the link to the DDRH Lagrangian and RMF theory is
evident: Evaluating eq.(\ref{Lagrangian}) in DHF approximation
self-energies of the same structure are obtained.

A self-contained model is obtained by introducing an in-medium
"renormalization" scheme. For that purpose it is of advantage to
consider symmetric hyper-matter, i.e. $k^N_F=k^Y_F=k_F$ where one
finds the relation
\beq\label{YNvertex} \Gamma_{\alpha Y}(k^Y_F)=\Gamma_{\alpha
N}(k^Y_F)\frac{\Sigma^Y_\alpha(k|k_F)}
{\Sigma^N_\alpha(k|k_F)}\vert_{k=k_F,k^N_F=k^Y_F}.
\eeq
which is exact in Hartree approximation. The significance of this
result becomes apparent by expanding DB self-energies
diagramatically with respect to the bare coupling constants
$g^B_\alpha$ \cite{KHL00}. Because the leading order contributions
are given by tadpole diagrams one derives easily that the nucleon
and hyperon self-energies are related by scaling laws
\beq\label{RY}
R^Y_\alpha=\frac{\Sigma^Y_\alpha}{\Sigma^N_\alpha}\simeq
\frac{g^Y_\alpha}{g^N_\alpha} (1 +{\cal
O}(1-\frac{M_N}{M_Y}))+\cdots
\eeq
where the realistic case $g^Y_\alpha < g^N_\alpha$ is considered.
The scaling factors $R^Y_\alpha$ are expected to be
state-independent, universal constants whose values are close to
the ratios of the bare coupling constants. For asymmetric
hypermatter with a hyperon fraction $\zeta_Y=\frac{\rho^Y}{\rho^N}
\ll 1$ a corresponding diagrammatic analysis shows that asymmetry
terms are in fact suppressed because the asymmetry correction is
of leading second order ${\cal O}((\frac{g^Y}{g^N}\zeta_Y)^2)$.
Thus, even in a finite nucleus where $\zeta_Y$ may vary over the
nuclear volume, we expect $R^Y_\alpha=const.$ to a very good
approximation.

These results agree remarkably well with the conclusions drawn
from the analyis of single hypernuclei in purely phenomenological
models. In the present context, eqs.(\ref{YNvertex}) and
(\ref{RY}) are of particular interest because they allow to
extend the DDRH approach in a theoretically meaningful way to
hypernuclei using the results available already from
investigations of systems without strangeness. Below, the nucleon
(Hartree) scalar and vector vertex functions
$\Gamma_{\sigma,\omega N}(k_F)$ of \cite{FL95} will be used as
reference values. The hyperon scaling factors $R^Y_\alpha$
are treated as phenomenological constants to be determined
empirically. The $\chi^2$ distribution from fitting DDRH
RMF-calculations to existing hypernuclear data deduced from
$(\pi,K)$ \cite{Ha96,Aj95,Pi91,Da86,Ma97} experiments is shown in
Fig.~1. The joined distribution of the scalar and vector scaling
factors is characterized by a steep valley along the diagonal
without a clearly pronounced minimum.

 \begin{figure}
 \label{chisquare}
 \centering\epsfig{file=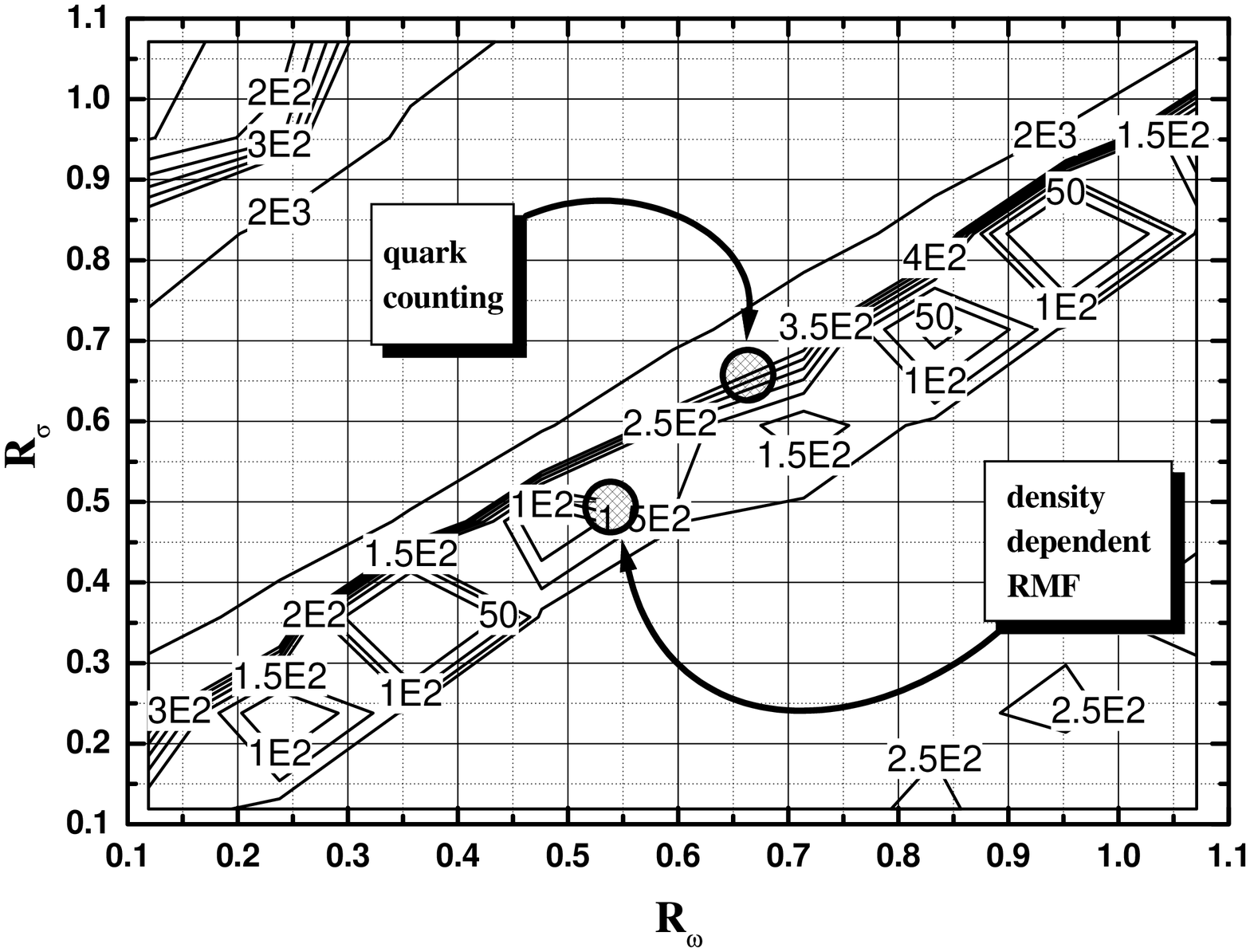,width=10cm,height=6cm}
 \caption{$\chi^2$ distribution for the scalar (R$_\sigma$) and vector
 (R$_\omega$) scaling factors. DDRH results for
 variations of $(R_\sigma, R_\omega)$ are compared
 to $\Lambda$ single particle spectra obtained from $(\pi^+,K^+)$
 reactions \protect\cite{Ha96,Aj95,Pi91,Da86,Ma97}. The location of
 DDRH coupling constants and the values assumed in the naive SU(3)$_f$ quark
 counting model are indicated.}
 \end{figure}
 \pagebreak

In order to stay as close as possible to the microscopic DDRH
picture we use the theoretically derived value R$_\sigma$ = 0.490
obtained from free space N$\Lambda$ scattering with the J\"ulich
potential \cite{Ha98}. Since theoretical values for the $\omega$
vertex are not available $R_\omega$ is taken from Fig.~1 leading
to R$_\omega$=0.553 with the above value of R$_\sigma$
\cite{KHL00}. The scaling factors are in surprisingly good
agreement with RMF results \cite{Ma96} and are consistent with
bounds on hyperon--nucleon couplings extracted from neutron star
models \cite{Gl92,Hu98}.


\section{Spectroscopy of single $\Lambda$ hypernuclei}
\label{sec:Spectroscopy}

Relativistic DDRH Hartree theory and applications to isospin
nuclei were discussed in great detail in ref.~\cite{FL95} and the
references therein. Here, we present DDRH results only for single
$\Lambda$ hypernuclei. The vertices are taken from DB calculations
with the Bonn A NN potential \cite{BM90}. The model parameters are
compiled in Tab.~I.

\subsection{$\Lambda$ Single particle states and spin-orbit splitting}
\label{ssec:SPStates}

Hyperon single particle spectra for $S$=-1 hypernuclei can be seen
as a very clean fingerprint of mean-field dynamics, since they
are only weakly affected by many-body effects. The bulk structure
contains information on the mean-field and, indirectly, the
nucleonic density distributions. Of particular theoretical
interest are spin-orbit splittings. At high energy resolution the
fine structure of the spectra provides information also on
dynamical correlations beyond static mean-field dynamics.

DDRH $\Lambda$ single particle levels for light to heavy nuclei
are shown in Fig.~2. Two major differences between neutron and
$\Lambda$ spectra are observed \cite{KHL00}:
\begin{enumerate}
\item $\Lambda$ and neutron single particle spectra are overall related
by a constant shift and an additional compression because the
$\Lambda$ central potential has a depth of only about -30 MeV,
compared to -70 MeV for the neutrons.
\item The spin-orbit splitting of the $\Lambda$ states is reduced further
being less than what is expected from the overall reduction of the
potential strength.
\end{enumerate}
 \begin{figure}
 \label{spectra}
 \centering\epsfig{file=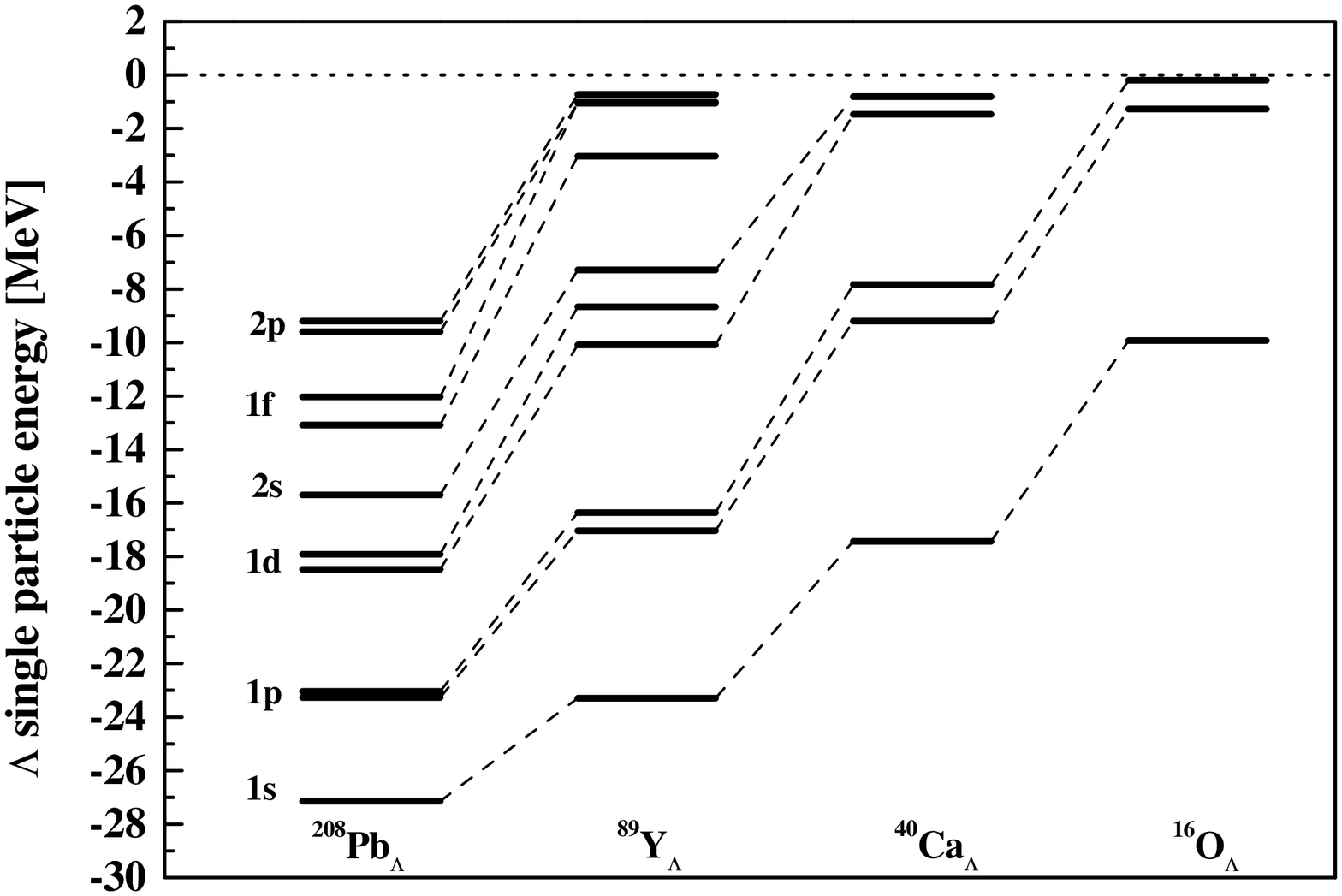,width=10cm,height=6cm}
 \caption{DDRH $\Lambda$ single particle spectra for light to heavy nuclei.}
 \end{figure}
The DDRH calculations reproduce the experimentally observed very
small spin-orbit splitting in $\Lambda$ hypernuclei,
e.g.~\cite{Ma81,Sa99}, reasonably well even without an explicit
dynamical suppression of spin-orbit interactions as e.g. a
$\Lambda-\omega$ tensor coupling which, for example, is used in
the QMC model \cite{Ts98}. $\Lambda$ and neutron spin-orbit
potentials are compared in Fig.~3.
\begin{figure}
\centering\epsfig{file=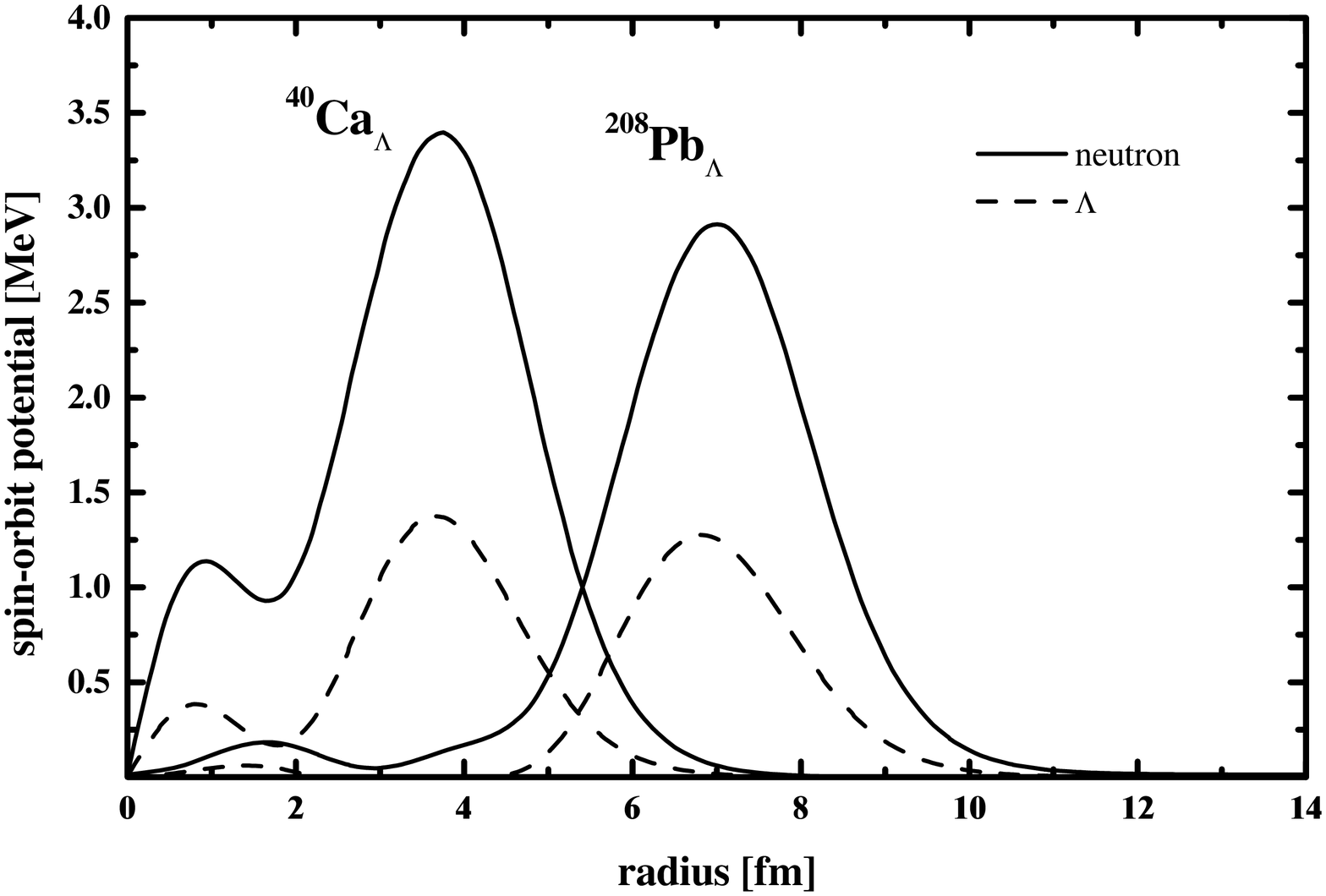,width=10cm,height=6cm}
\caption{$\Lambda$ and neutron spin-orbit potentials in
$^{40}$Ca$_\Lambda$ and $^{208}$Pb$_\Lambda$.} \label{so-pot}
\end{figure}
The reason for the small spin-orbit splitting is understood by
considering the evolution with increasing mass number. From Fig.~4
it is seen that the splitting drops for higher masses. Such a
behaviour is to be expected since the spin-orbit potential is a
finite size effect and will vanish in the nuclear matter limit. A
corresponding mass dependence is also found in pure isospin
nuclei. The splitting also drops in the low mass region -- now
for the reason that the spin-orbit doublets approach the
continuum threshold and get compressed before one of them or both
become unbound. There is a remarkable similarity to the situation
found in weakly bound neutron-rich exotic nuclei \cite{Le98}. In
both cases weak binding is an important reason for the reduction
of spin-orbit effects. The wave functions of weakly bound states
- either for the special dripline neutron states or $\Lambda$
orbits in general - have a reduced overlap with the spin-orbit
potential which remains well localized in the nuclear surface. As
seen from Tab.~II the binding energy effect leads to substantially
larger spatial extensions of the $\Lambda$ states.
  \begin{figure}
 \label{fig:so-split}
 \centering\epsfig{file=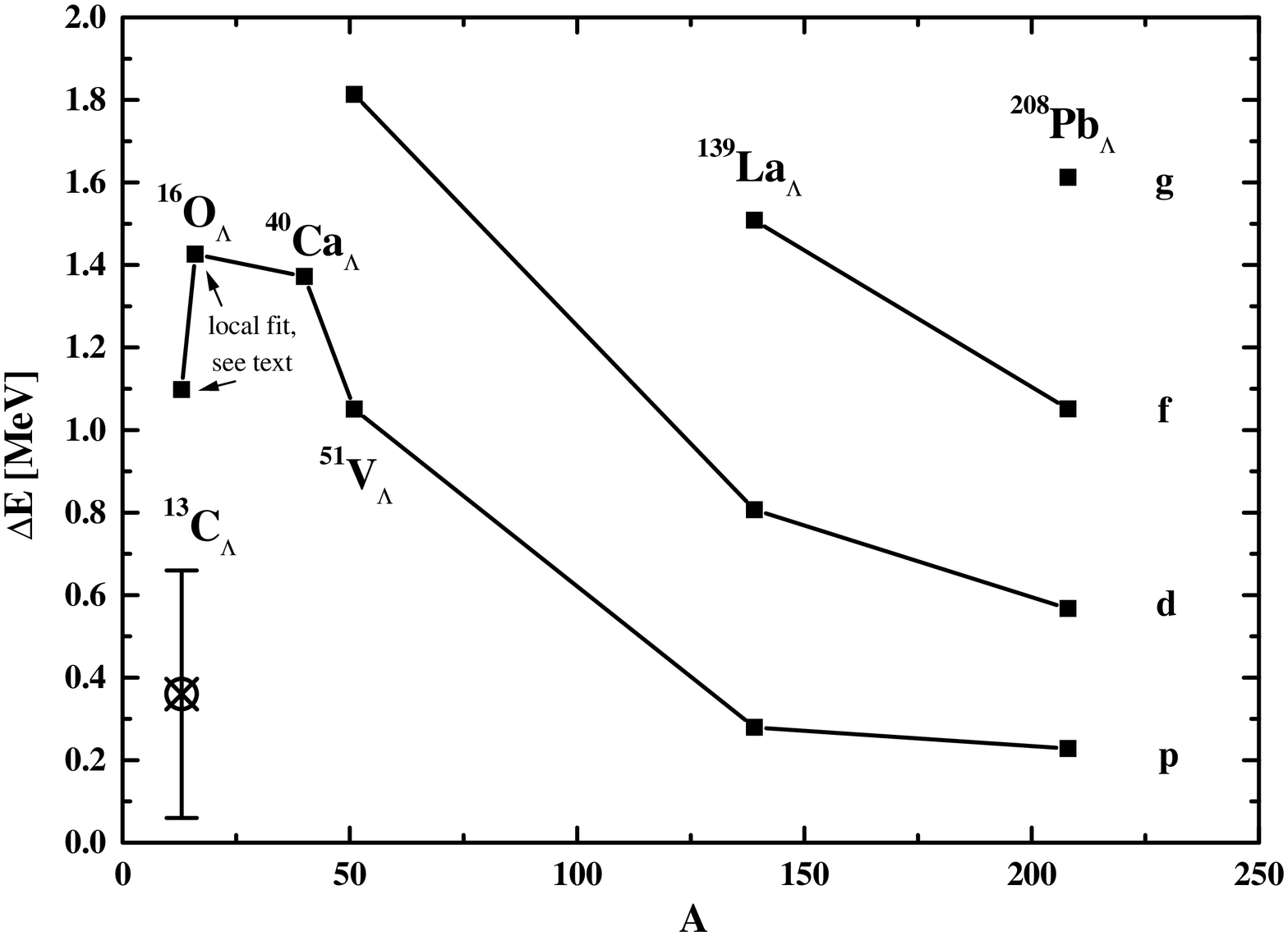,width=10cm,height=6cm}
 \caption{Spin-orbit splitting of single $\Lambda$ levels.
For our standard choice of coupling constants the $^{13}$C data
point \protect\cite{Ma81} is overestimated but reducing the
vector coupling by about 2\% (R$_\omega$=0.542) a better description is
obtained (see text).}
 \end{figure}
%

\subsection{Comparison to data and phenomenological RMF calculations}
\label{ssec:DDvs}

In Fig.~5 the DDRH single particle spectra are compared to
spectroscopic data from $(\pi^+, K^+)$ reactions. States in
intermediate to high mass nuclei are described fairly well by the
model, while for masses below about $^{28}Si_\Lambda$ deviations
of up to 2.5 MeV arise, possibly indicating the limits of an
approach using nuclear matter vertices in local density
approximation. We seem to miss sytematically a decrease of the
repulsive vector interaction for low mass single $\Lambda$
hypernuclei. This tendency already becomes apparent going from
$^{51}V_\Lambda$ to $^{28}Si_\Lambda$. Actually, the dashed line
for $R_\omega = 0.542$ in Fig.~5 shows that a slight reduction of
the vector repulsion improves considerably the description of the
light mass data. This 2\% variation of $R_\omega$ is largely
within the uncertainties of the model parameters.
\begin{figure}
\centering\epsfig{file=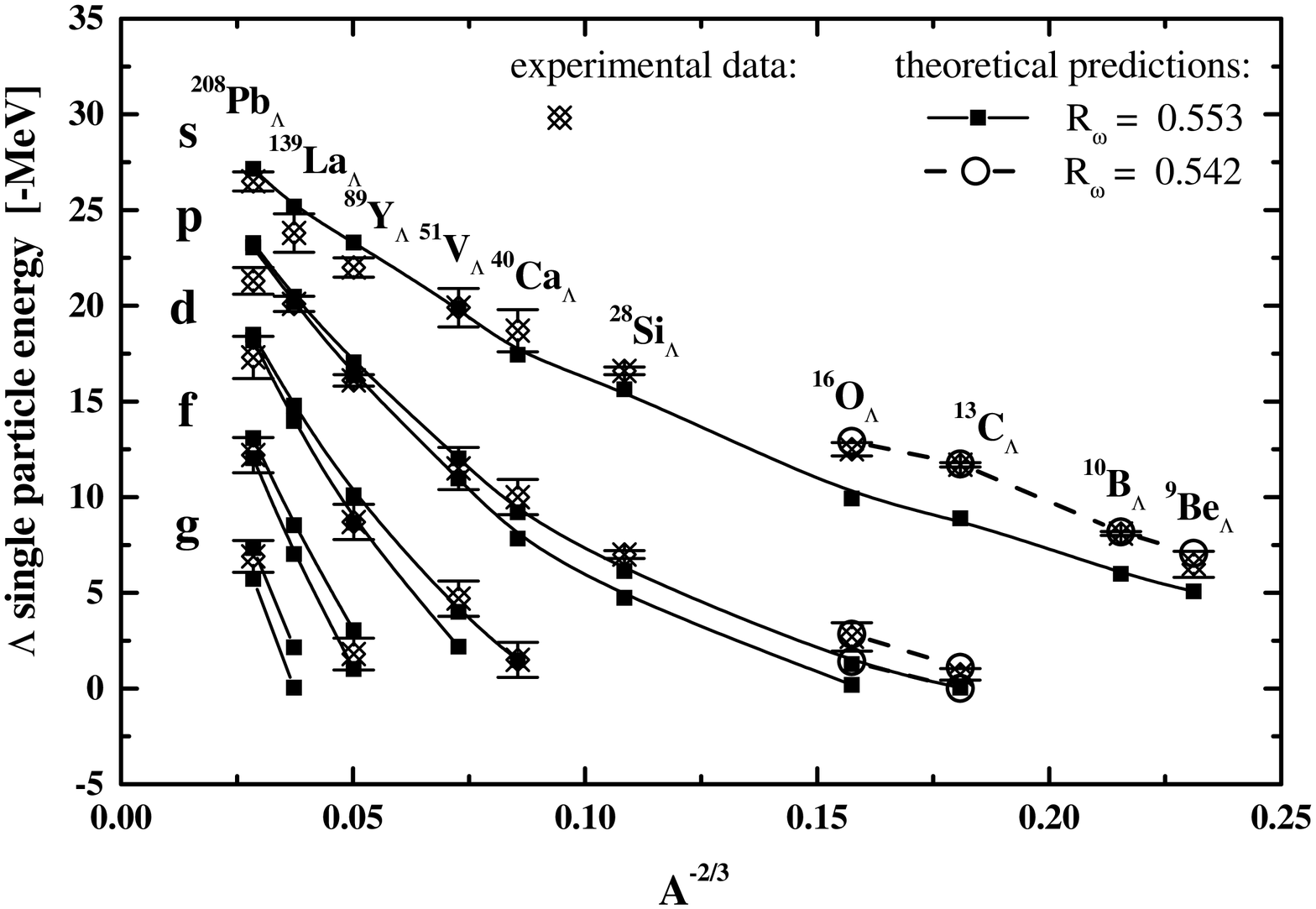,width=10cm,height=6cm}
 \caption{Comparison of DDRH single $\Lambda$ separation energies
 to data from
$^AX (\pi^+, K^+) ^AX_\Lambda$ reactions
\protect\cite{Ha96,Aj95,Pi91,Da86,Ma97}.} \label{separation}
 \end{figure}
Especially for the heavier nuclei the microscopic DDRH results
are at least of comparable quality as the phenomenological
descriptions. This we consider as a remarkable success of the
model because the coupling functionals were not especially
adjusted to data except for the overall adjustment of the vector
scaling factor $R_\omega$. In heavy nuclei the density dependence
of the $\Lambda$ vertices act mainly as an overall reduction
factor because the $\Lambda$ vector density is relatively small
and varies only weakly across the nuclear volume. The essential
density dependent effect is related to the dynamical
redistribution of the surrounding nucleons by the rearrangement
self-energies. Dynamically, it corresponds to a modification of
the $\Lambda$ core potential due to static polarization in the
nucleonic sector. Obviously, this effect is not accounted for by
conventional RMF models.

Finally, DDRH and RMF results for $\Lambda$ particle-neutron hole
configurations observed in ($K^-,\pi^-$) reactions
\cite{Po80,Be79} are given in Tab.~III.

\section{Summary, conclusions and outlook}
\label{sec:summary}

The DDRH theory introduced previously for
isospin nuclei was extended to hypernuclei by including the full
set of SU(3)$_f$ octet baryons. Interactions were described by a
model Lagrangian including strangeness-neutral scalar and vector
meson fields of $q\overline{q}$ (q=u,d) and $s\overline{s}$ quark
character. The medium dependence of interactions was described by
meson-baryon vertices chosen as functionals of the baryon field
operators. The DDRH vertices are chosen to cancel Dirac-Brueckner
ground state correlations. Hence, the approach corresponds to a
resummation of ladder diagrams into the vertices under the
constraint that infinite matter ground state self-energies and
total binding energies are reproduced. As the central theoretical
result it was found that the structure of Dirac-Brueckner
interactions strongly indicates that the ratio of nucleon and
hyperon in-medium vertices should be determined already by the
ratio of the corresponding free space coupling constants being
affected only weakly by the background medium.

Dynamical scaling of nucleon-hyperon vertices was tested in DDRH
mean-field calculations for single $\Lambda$ hypernuclei.
Calculations over the full range of known single $\Lambda$ nuclei
led to a very satisfactory description of $\Lambda$ separation
energies. The deviations from the overall agreement for masses
below A$\approx$16 are probably related to the enhancement of
surface effects in light nuclei which are not described properly
by static RMF calculations with DB vertices obtained in the local
density approximation. In a recent non-relativistic calculation
indeed sizable contributions of hyperon polarization
self-energies especially in light nuclei \cite{Vi98} were found.

The results are encouraging and we conclude that DDRH theory is in
fact an appropriate basis for a microscopic treatment of
hypernuclei. The present formulation and applications are first
steps on the way to a more general theory of in-medium SU(3)$_f$
flavor dynamics. Future progress on dynamical scaling and other
theoretical aspects of the approach is depending on the
availability of Dirac-Brueckner calculations for the full baryon
octet including also the complete pseudoscalar $0^-$ and vector
$1^+$ meson multiplets.

As work in progress the production of hypernuclei in hadronic
reactions is presently investigated. Initial and final state
interactions of the incident and outgoing mesons in
($\pi^+,K^+$) and ($K^-,\pi^-$) reactions are described in a
relativistic eikonal approach. Good agreement with differential
and total cross section data is obtained \cite{Briganti00}. A
Lagrangian model, including s-channel production through nucleon
resonances and t-channel production by e.g. $K^*$ exchange is
under investigation. The nuclear structure results discussed here
are entering into the (coherent) production amplitude.
Electro-production of hypernuclei will be used as an independent
and important test for the production vertex and the nuclear
structure input.

\section*{Acknowledgements}

This work was supported in part by DFG under contract Le439/4-3,
Graduiertenkolleg Theoretische und Experimentelle
Schwerionenphysik, GSI and BMBF. Discussions with C. Greiner are
gratefully acknowledged.


%



%
\begin{table}
\label{tab:ParaTab} \caption{Hadron masses and DDRH coupling
constants at saturation density $\rho_0$=0.16~fm$^{-3}$ for
$\Lambda$ particles and nucleons.}
\begin{tabular}{c|c||c|c|c}
 & Mass & Coupling Constants (at $\rho_0$) & $B$ = N & $B$ = $\Lambda$ \\ \hline \hline
 $m_N$       &      939.0 MeV & $g_{\sigma B}^2/4\pi$      &
6.781 &  1.628 \\
 $m_\Lambda$ &1115.0 MeV & $g_{\omega B}^2/4\pi$ &
9.899 &  3.022 \\
 $m_\sigma$  & 550.0 MeV & $g_{\rho B}^2/4\pi$   &
1.298 & 0.000 \\
 $m_\omega$ & 782.6 MeV & & & \\
 $m_\rho$    & 770.0 MeV & & \\
\end{tabular}
\end{table}
\vspace{3mm}
 \begin{table}
 \label{rms-radii}
 \caption{r.m.s. radii for $\Lambda$,
neutron and proton single particle states in $^{40}Ca_\Lambda$ and
$^{208}Pb_\Lambda$}
 \begin{tabular}{l|ccc|ccc}
    & \multicolumn{3}{c}{$^{40}Ca_\Lambda$}
    & \multicolumn{3}{c}{$^{208}Pb_\Lambda$} \\
    & $\Lambda$ & $n$ & $p$ & $\Lambda$ & $n$ & $p$ \\ \hline
    1s$_{1/2}$ & 2.8 fm & 2.3 fm & 2.4 fm &
    4.1 fm & 3.8 fm & 3.9 fm \\
    1p$_{3/2}$ & 3.5 fm & 3.0 fm & 3.0 fm &
    4.8 fm & 4.5 fm & 4.6 fm \\
    1p$_{1/2}$ & 3.6 fm & 3.0 fm & 3.0 fm &
    4.7 fm & 4.4 fm & 4.5 fm \\
    1d$_{5/2}$ & 4.7 fm & 3.5 fm & 3.6 fm &
    5.3 fm & 5.0 fm & 5.1 fm \\
    1d$_{3/2}$ & 6.3 fm & 3.6 fm & 3.7 fm &
    5.2 fm & 4.9 fm & 5.0 fm \\
 \end{tabular}
 \end{table}
\vspace{3mm}
 \begin{table}
 \label{ph-states}
 \caption{Transition energies of $\Lambda$ particle--neutron hole
 excitations in single
$\Lambda$ hypernuclei observed in $(K^-,\pi^-)$ reactions. DDRH
results and phenomenological RMF calculations \protect\cite{Ru90}
(phen. RMF), including nonlinear $\sigma$ self interactions, are
compared to experimental values (exp.) \protect\cite{Po80,Be79}.}
 \begin{tabular}{l|l|l|d|d|d}
    & & & exp. & DDRH & phen.RMF \\
    & neutron valence state & configuration & [MeV] & [MeV] & [MeV] \\ \hline
    $^{12}C_\Lambda$ & $1p_{3/2}$ &
    $(1s_{1/2}\Lambda,1p_{3/2}n^{-1})$ &
    6.72$\pm$2 & 6.69 & 5.02 \\
    & & $(1p_{3/2}\Lambda,1p_{3/2}n^{-1})$ &
    18.48$\pm$2 & 15.11 & 17.21 \\ \hline
    $^{16}O_\Lambda$ & $1p_{1/2}$ &
    $(1s_{1/2}\Lambda,1p_{1/2}n^{-1})$ &
    3.35$\pm$2 & 5.76 & 3.53 \\
    & & $(1s_{1/2}\Lambda,1p_{3/2}n^{-1})$ &
    9.90$\pm$2 & 10.13 & 9.46 \\
    & & $(1p_{1/2}\Lambda,1p_{1/2}n^{-1})$ &
    13.20$\pm$2 & 16.16 & 13.89 \\
    & & $(1p_{3/2}\Lambda,1p_{3/2}n^{-1})$ &
    19.20$\pm$2 & 18.40 & 18.88 \\ \hline
    $^{40}Ca_\Lambda$ & $1d_{3/2}$ &
    $(1p_{1/2}\Lambda,1d_{3/2}n^{-1})$ &
    5.79$\pm$2 & 8.84 & 7.40 \\
    & & $(1d_{3/2}\Lambda,1d_{3/2}n^{-1})$ &
    14.47$\pm$2 & 11.34 & 15.48 \\
    & & $(1d_{5/2}\Lambda,1d_{5/2}n^{-1})$ &
    19.35$\pm$2 & 20.07 & 20.71
 \end{tabular}
 \end{table}

\end{document}